\documentclass[adp,fleqn]{w-art}
\usepackage{times}
\usepackage{w-thm}
\usepackage{bm} 
\usepackage[]{graphicx}
\chardef\bslash=`\\ 

\begin{document}
\DOIsuffix{theDOIsuffix}
\Volume{12}
\Issue{1}
\Copyrightissue{01}
\Month{01}
\Year{2003}
\pagespan{1}{}
\keywords{high--energy hadron--hadron collisions, 
generalized parton distributions, quantum chromodynamics}
\subjclass[pacs]{12.38.-t, 13.85.-t, 14.80.Bn} 
%
%
\title[3D parton imaging]{3D parton imaging of the nucleon
in high--energy $pp$ and $pA$ collisions}
\author[Frankfurt]{L.~Frankfurt}
\address[\inst{1}]{School of Physics and Astronomy, Tel Aviv University,
Tel Aviv, Israel}
\author[Strikman]{M.~Strikman}
\address[\inst{2}]{Department of Physics, Pennsylvania State University,
University Park, PA 16802, U.S.A.}
\author[Weiss]{C.~Weiss}
\address[\inst{3}]{Theory Group, Jefferson Lab, Newport News, 
VA 23606, U.S.A.}
\dedicatory{Dedicated to Klaus Goeke on the occasion of his 60th birthday}
\begin{abstract}
We discuss several examples of how the transverse spatial distribution
of partons in the nucleon, as well as multiparton correlations, can be
probed by observing hard processes (dijets) in high--energy
$pp \; (\bar p p)$ and $pA \; (dA)$ collisions. Such studies 
can complement the information gained from measurements of hard exclusive
processes in $ep$ scattering. The transverse spatial distribution of partons 
determines the distribution over $pp$ impact parameters of events with 
hard dijet production. Correlations in the transverse positions of partons
can be studied in multiple dijet production. We find that the correlation
cross section measured by the CDF Collaboration, 
$\sigma_{\rm eff} = 14.5\pm 1.7^{+ 1.7}_{-2.3} \; {\rm mb}$,
can be explained by ``constituent quark'' type quark--gluon 
correlations with $r_q \approx r_N / 3$, as suggested by the 
instanton liquid model of the QCD vacuum. Longitudinal 
and transverse multiparton correlations can be separated in a 
model--independent way by comparing multiple dijet production 
in $pp$ and $pA$ collisions. Finally, we estimate the cross section
for exclusive diffractive Higgs production in $pp$ collisions at LHC
(rapidity gap survival probability), by combining the impact parameter 
distribution implied by the hard partonic process with information 
about soft interactions gained in $pp$ elastic scattering. 
\end{abstract}
\maketitle                   
\section{Introduction}
Hard exclusive processes in $ep$ scattering allow to probe not only
the distribution of partons with respect to longitudinal momentum,
but also their spatial distribution in the transverse plane.
Examples include the hard electroproduction of light mesons 
and real photons (deeply virtual Compton scattering), as well as
the photoproduction of heavy quarkonia ($J/\psi, \psi^{\prime},\Upsilon$).
Thanks to QCD factorization theorems the amplitudes for these processes 
can be separated into a ``hard'' part, calculable in perturbative QCD, 
and ``soft'' parts characterizing the non-perturbative structure of 
the involved hadrons. The information about the nucleon is contained
in so-called generalized parton distributions (GPD's). These are functions
of the parton momentum fractions, $x$ and $x'$, as well as 
of the invariant momentum transfer to the nucleon, $t$, and thus
combine aspects of the usual parton distributions, measured
in inclusive deep--inelastic scattering, with those
of the elastic nucleon form factors. For $x = x'$, their Fourier transform 
with respect to $t$ describes the spatial distribution of partons
in the transverse plane. The GPD's thus, in a sense, 
provide us with a ``3D parton image'' of the nucleon.

Exclusive processes in $ep$ scattering, however, are not the
only reactions which probe the ``3D'' parton distributions.
In fact, a lot more information about the longitudinal momentum
and transverse spatial distribution of partons, as well as 
about multiparton correlations, can be obtained from the study
of selected hard processes in (not necessarily exclusive) 
$pp$ and $pA$ scattering. Comparative studies of $ep$ and $pp/pA$ 
induced hard processes will help to improve the quantitative description 
of both classes of processes and offer many new, fascinating insights 
into the partonic structure of the nucleon. The conceptual basis for 
such a program is the combination of Gribov's
space--time picture of hadron interactions at high energies, which allows 
for a unified description of $ep$ and $pp$ scattering \cite{Gribov:jg},
and the QCD factorization theorems for hard exclusive processes.
In view of the planned new experiments in both $ep$ scattering
(Jefferson Lab at 12 GeV, EIC/eRHIC) and $pp / pA$ scattering (RHIC, LHC),
such studies are very timely.

In this paper we study various examples of hard processes 
in $pp$ and $pA$ scattering which probe the ``3D distribution''
of partons and their correlations in the nucleon. First, we 
show how the production of multiple dijets in $pp$ events
can resolve spatial correlations of partons in the transverse plane.
These correlations provide interesting new information about the
long--wavelength structure of the nucleon, which can {\it e.g.}\
be related to the role of instanton--type vacuum fluctuations
in generating constituent quark masses (spontaneous breaking of 
chiral symmetry), which is the basis for the chiral quark--soliton model 
of the nucleon \cite{Diakonov:2002fq}. Second, we discuss how similar 
experiments with nuclear targets ($pA, dA$) can help to separate between 
longitudinal and transverse parton--parton correlations in the nucleon 
wave function. Third, we discuss the role of the transverse spatial 
distribution of partons in the diffractive production of heavy particles
(Higgs bosons) in $pp$ collisions at LHC energies. Such diffractive
events involve a delicate interplay of ``hard'' and ``soft'' processes
(survival of the rapidity gaps), and the latter can be modeled relying 
on information gained from $pp$ elastic scattering.
\section{The transverse spatial distribution of partons
and hard processes in $pp$ scattering}
\label{sec_distribution}
In order to define the transverse spatial distribution of 
gluons in the nucleon, it is convenient to write the 
gluon GPD in the form\footnote{We are considering here the GPD for 
zero longitudinal momentum transfer; in the general case the 
GPD would depend on the momentum fractions of the two gluons
separately, $x' \neq x$.} (analogous expressions apply to
the quark flavor singlet and non-singlet distributions)
\begin{equation}
H_g (x, t, Q^2) \;\; = \;\; g(x, Q^2) \; F_g (x, t, Q^2) ,
\end{equation}
where $g(x,Q^2) = H_g (x, t = 0,Q^2)$ is the usual gluon density,
and $F_g (x, t,Q^2)$ the ``two-gluon form factor'' of the nucleon,
$F_g (x, t = 0,Q^2) = 1$. In the following, the $Q^2$--dependence of 
the form factor and related quantities will not be indicated explicitly. 
One can represent this form factor as the Fourier transform 
of a function of a transverse coordinate variable, $\bm{\rho}$, 
\begin{equation}
F_g (x, t ) \;\; = \;\; \int d^2 \rho \; 
e^{i (\bm{\Delta}_\perp \bm{\rho})} \; F_g (x, \rho) ,
\hspace{4em}
(t = -\bm{\Delta}_\perp^2 ) .
\label{impact_def}
\end{equation}
cf.\ the well--known relation between the nucleon electric form factor
and the charge density in the Breit frame. The function $F_g (x, \rho)$ 
describes the spatial distribution of gluons with longitudinal momentum 
fraction $x$ in the transverse plane, with 
$\int d^2 \rho \, F_g (x, \rho) = 1$. A measure of the 
gluonic transverse size of the nucleon for given $x$ is the average
\begin{equation}
\langle \rho^2 \rangle
\;\; \equiv \;\; \int d^2 \rho \; \rho^2 \; F_g(x, \rho)
\;\; = \;\; 4 \; \frac{\partial}{\partial t}
F_g (x, t)_{t = 0} . 
\end{equation}

On general grounds, the gluonic transverse size is expected
to grow with decreasing $x$. Different physical mechanisms are responsible 
for this growth in different regions of $x$. Near $x = 1$, 
the growth of $\langle \rho^2 \rangle$ is governed by the 
Feynman mechanism --- the $t$--dependence of the two--gluon form factor
disappears if the active gluon carries the entire longitudinal 
momentum of the nucleon, $x \rightarrow 1$, causing $\langle \rho^2 \rangle$ 
to vanishes at $x = 1$. Note that this behavior is a trivial consequence
of the relativistic kinematics of the form factor calculation;
it does not reflect the transverse position of the $x \rightarrow 1$ 
parton relative to the transverse region occupied by the spectator
system. Hence, in a proton-proton collision (see below) there
is no simple relation between $F_g (x, \rho)$ and the $pp$
impact parameter in the $x$--region dominated by the Feynman mechanism.
When $x$ is decreased below the valence region, 
a distinctive increase of $\langle \rho^2 \rangle$ is caused by 
pion cloud contributions to the gluon density, which set in for 
$x < M_\pi / M_N$ \cite{Strikman:2003gz,Frankfurt:2002ka}.
Finally, when $x$ is decreased further, the transverse size grows
due to the random walk character of successive emissions 
in the partonic ladder (Gribov diffusion) \cite{Gribov:jg}.
\footnote{For a cross sections increasing as a power of the energy, 
unitarity at small impact parameters also leads to an increase of the 
radius of the interaction with energy.}

As to the $Q^2$ dependence, the nucleon's gluonic transverse size at 
fixed $x$ decreases with increasing $Q^2$ as a result of DGLAP evolution;
see Ref.~\cite{Frankfurt:2003td} for a discussion of this effect.

In $ep$ scattering, the cleanest way to probe the transverse spatial
distribution of gluons is via exclusive photo-- or electroproduction of
heavy quarkonia ($J/\psi, \Upsilon$). These processes have been measured 
in a number of fixed--target experiments ($x \geq 10^{-1}$), as well
as in the H1 and ZEUS experiments at the HERA collider
($x \sim 10^{-2} - 10^{-3}$), see Ref.~\cite{DIS04} 
for a recent review of the data. The two--gluon form factor can 
be extracted from the measured $t$--dependence of the differential 
cross section. It was found that $\langle \rho^2 \rangle$ increases 
from $\sim 0.24 \, {\rm fm}^2$ at $x \sim 10^{-1}$ to 
$\sim 0.35 \, {\rm fm}^2$ at $x \sim 10^{-2} - 10^{-3}$;
this difference can quantitatively be explained by the
pion cloud contribution which is switched on for $x < M_\pi / M_N$ 
\cite{Strikman:2003gz}. Concerning the shape of the two-gluon form factor, 
it has been argued that at $x \geq 10^{-1}$ (where pion cloud contributions
are absent) the two--gluon form factor should follow the axial form factor
of the nucleon, and thus be described by the dipole parametrization
\begin{equation}
F_g (x, t) \;\; = \;\; (1 - t/m_g^2)^{-2},
\hspace{4em} m_g^2 \;\; = \;\; 1.1 \, {\rm GeV}^2 
\hspace{4em} (x \geq 10^{-1}).
\label{dipole}
\end{equation}
This form indeed describes well the $t$--dependence of the fixed--target
$J/\psi$ photoproduction experiments. Based on Eq.~(\ref{dipole}), 
we have suggested in Ref.~\cite{Frankfurt:2003td} a generalized
dipole parametrization valid also at small $x$, which incorporates
the observed increase in the gluonic transverse size (as well as the
effects of DGLAP evolution) by way of an $x$-- and $Q^2$--dependent
of the dipole mass parameter, $m_g^2 (x, Q^2)$. This simple parametrization
conveniently summarizes our knowledge of the transverse spatial 
distribution of gluons from $ep$ scattering. For details, see
Refs.~\cite{Frankfurt:2003td,DIS04}.

The transverse spatial distribution of quarks in the nucleon can be probed 
in the production of neutral vector mesons at sufficiently large $x$,
and also in processes where gluon exchange does not contribute, 
such as the production of charged vector mesons like
$\gamma^*+ p\rightarrow \rho^+ + n$. 

Turning now to $pp$ collisions, an immediate application
of the transverse spatial distribution of partons is in
the description of the impact parameter dependence of the
cross section for hard dijet production \cite{Frankfurt:2003td}.
In a $pp$ collision with c.m.\ energy $\sqrt{s}$, a hard dijet with 
transverse momentum $q_\perp$ at zero rapidity is produced 
in the collision of two partons carrying momentum fractions
\begin{equation}
x_1 = x_2 = 2 q_\perp / \sqrt{s}
\end{equation}
of the respective protons; the generalization to non-zero rapidity, 
$x_1 \neq x_2$, is straightforward. The probability for such a parton--parton 
collision, as function of the impact parameter of the $pp$ system, $b$,
is given by the convolution of the spatial distributions of the
partons; for gluons
\begin{equation}
P_2 (b) \;\; \equiv \;\; \int d^2\rho_1 \int d^2\rho_2 \; 
\delta^{(2)} (\bm{b} - \bm{\rho}_1 + \bm{\rho}_2 )
\; F_g (x_1, \rho_1 ) \; F_g (x_2, \rho_2 ) .
\label{P_2}
\end{equation}
The scale of the parton distributions here is $q_\perp^2$.
One can easily generalize this to the production of multiple dijets.
Neglecting possible correlations in the transverse positions
of two partons in the same proton (this question will be discussed
in detail in Section~\ref{sec_correlations}), 
the $b$--dependent probability for 
the production of a double dijet would be given by
\begin{equation}
P_4 (b) \;\; \equiv \;\; \frac{\left[ P_2 (b)\right]^2}
{\int d^2 b \; \left[ P_2 (b)\right]^2} .
\label{P_4}
\end{equation}

While exclusive processes in $ep$ scattering provide in principle 
the cleanest way to access the transverse spatial distribution
of partons, there are several instances in which $pp$ scattering
is more effective. One is the study of large $x$, where the
cross sections for exclusive processes in $ep$ are small. Besides, 
in this kinematics, when $x \sim 1$ are probed, one is mostly 
sensitive to the Feynman mechanism; see the discussion above.
In this case the global transverse distribution of matter can
be measured more directly using various reactions combining 
a soft and hard trigger, in particular in connection with 
$pA$ collisions \cite{Frankfurt:1985cv}. New opportunities 
for such studies will emerge at LHC, where the high luminosity will allow, 
for example, to compare the characteristics of $W^+$ and $W^-$ production 
at the same forward rapidities, corresponding to relatively high $x$ where
the $d/u$ ratio deviates strongly from the naive value 1/2. 
By studying the accompanying production of hadrons one can
learn which configurations in the nucleon have larger transverse size
--- those with a leading u--quark or with a leading d-quark. 
One suitable observable is, for example, the distribution of the 
number of events over the number of the produced soft particles. 
A larger transverse size corresponds to a larger probability 
of soft interactions, and hence to a larger probability of events 
with large multiplicity. It is interesting to note that the 
studies of the associated soft hadron multiplicity in the production 
of $W^{\pm}$ and $Z$ bosons in $\bar p p$ collisions by the 
CDF collaboration at Fermilab find an increase of this multiplicity 
by a factor of two as compared to generic inelastic 
events \cite{Field:2002vt}. This appears natural if one takes 
into account that the hard quarks producing the weak bosons have 
a narrower transverse spatial distribution than the soft partons.
As a result, the average impact parameters in events with weak boson
production are much smaller than in generic inelastic collisions, 
leading to an enhancement of multiple soft and semi-hard 
interactions \cite{Frankfurt:2003td}. 
\section{Probing correlations in the proton parton wave function 
via multiple dijet production}
\label{sec_correlations}
Single parton densities and GPDs do not carry information about
longitudinal and transverse correlations of partons in the hadron wave
function. Such information can be extracted from high energy $pp$ and
$pA$ collisions where two (or more) pairs of partons can collide to
produce multiple dijets, with a kinematics distinguishable from those
produced in $2 \rightarrow 4$ parton processes. Since the momentum scale 
of the hard interaction, $p_t$, corresponds to much smaller transverse 
distances in coordinate space than the hadronic radius, in a double
parton collision the two interaction regions are well separated in
transverse space. Experimentally, one measures the ratio
\begin{eqnarray}
\lefteqn{
\frac{\displaystyle
\frac{d\sigma}{d\Omega_{1}d\Omega_{2}d\Omega_{3}d\Omega_{4}}
(p+\bar p\to \mbox{jet1} + \mbox{jet2} + \mbox{jet3} + \gamma)}
{\displaystyle
\frac{d\sigma}{d\Omega_{1}d\Omega_{2}} (p+\bar p\to \mbox{jet1} 
+ \mbox{jet2} ) \cdot  
\frac{d\sigma}{d\Omega_{3}d\Omega_{4}} (p+\bar p\to \mbox{jet3} +\gamma) }} 
&&
\nonumber \\
&=& \frac{f(x_1,x_3,\mu^2)f(x_2,x_4,\mu^2)}
{\sigma_{\rm eff} \; f(x_1,\mu^2)f(x_2,\mu^2)f(x_3,\mu^2)f(x_4,\mu^2)} ,
\end{eqnarray}
where $f(x_1,x_3), f(x_2,x_4)$ are the longitudinal light-cone double parton 
densities at the hard scale $\mu^2$ (we assume for simplicity that 
the virtuality in both hard processes is comparable; in the following 
equations we suppress dependence on $\mu^2$), and the quantity 
$\sigma_{\rm eff}$ can be interpreted as the ``transverse correlation area''. 
The variables $\Omega_i$ characterize the observed jets (or photons) --- 
their transverse momenta, rapidities, cuts on the opening angle, etc.

Parton correlations can emerge due to nonperturbative effects
at a low resolution scale, or due to the effects of QCD evolution. 
One possible nonperturbative mechanism is the existence of
``constituent quarks'' within the nucleon, which appear due
to the interaction of current quarks with localized non-perturbative 
gluon fields, resulting in local short--range correlations 
in the transverse spatial distribution of gluons. The
instanton model of the QCD vacuum suggests a constituent quark
radius of about $1/3$ the nucleon radius, $r_q \approx  r_N / 3$.
Another nonperturbative mechanism, relevant at small $x$, are 
fluctuations of the color field in the nucleon due to the fluctuations 
of the transverse size of the quark distribution.
Perturbative correlations emerge due to small transverse distances
in the emission process in the perturbative partonic ladder in
DGLAP evolution. Of all the mentioned mechanisms, only the first one 
is effective at $x\ge 0.05$, where the data of the CDF experiment
were collected.
 
The CDF experiment observed correlation effects in a restricted 
$x$ range (two balanced jets, and jet plus photon) and found
$\sigma_{\rm eff} = 14.5\pm 1.7^{+ 1.7}_{-2.3} \; {\rm mb}$.
This value is significantly smaller than the naive estimate
obtained by taking a uniform distribution of partons of a
transverse size determined by the e.m.\ form factor of the nucleon,
which gives $\sigma_{\rm eff} \approx 53 \; {\rm mb}$, indicating 
strong correlations between the transverse positions of partons 
in the transverse plane. The longitudinal correlation between partons 
in the measured kinematics due to energy conservation is likely to be small,
as $x_1 + x_2$ and $x_3 + x_4$ are much smaller than 1. If this effect 
were important it would likely lead to a suppression of the double 
parton collision cross section, and hence to an increase
of $\sigma_{\rm eff}$. However, no dependence of $\sigma_{\rm eff}$ 
on $x_i$ was observed in the experiment.

For a more quantitative analysis of the CDF data, we can make use
of the information about the transverse spatial distribution of
gluons gained from $J/\psi$ photoproduction, as summarized in
Section~\ref{sec_distribution}. Since the $x$ values of the partons 
probed were reasonably small compared to 1, the simple ``geometric''
picture of the $\bar p p$ collision in transverse position 
in the spirit of Eqs.~(\ref{P_2}) and (\ref{P_4}) is justified,
and one has
\begin{equation}
\sigma_{\rm eff} \;\; = \;\; 
\left[ \int d^2 b \; P_2^2(b) \right]^{-1} .
\end{equation}
Evaluating this with the dipole parametrization of the 
two--gluon form factor (\ref{dipole}), this comes to
\begin{equation}
\sigma_{\rm eff} \;\; = \;\; 
\frac{28\pi }{m_g^2}
\;\; \approx \;\; 34\; \mbox{mb} . 
\end{equation}
Thus, about 50\% of the enhancement compared to the naive estimate
of the previous paragraph is due to smaller actual transverse radius
of the gluon distribution. Still, our value indicates significant
correlations in the transverse positions of the partons.
In the kinematics discussed here the relevant partons are both 
quarks and gluons. We can estimate the effect of correlations
assuming that most of the partons are concentrated in a small 
transverse area associated with the ``constituent quarks'',
as implied by the instanton liquid model of Diakonov and 
Petrov \cite{Diakonov:2002fq}. Assuming a constituent quark
radius of $r_q \sim r_N/3$, we obtain an enhancement factor
due to transverse spatial correlations of partons of
\begin{equation}
\frac{8}{9} + \frac{1}{9}\; \frac{r_N^2}{r_q^2} \;\; \sim \;\; 
1.6 \div 2 .
\end{equation}
This is roughly the value needed to explain the remaining discrepancy
with the CDF data. Thus, the combination of the relatively small 
transverse size of the distribution of large--$x$ gluons
and the quark--gluon correlations implied by ``constituent quarks''
with $r_q \approx r_N / 3$ is sufficient to explain the trend of 
the CDF data. Further studies of multijet events at hadron--hadron colliders, 
with a broader range of final states, would in principle allow to measure 
separately quark--quark, quark--gluon, and gluon--gluon correlations 
for different $x$.
 
However, studies based on $\bar pp$ or $pp$ collisions alone
do not allow for a model--independent separation of transverse and 
longitudinal correlations. This is possible only in $pA$ collisions
at RHIC and LHC.  The reason is that the nucleus, having a thickness 
which practically does not change on the nucleon transverse scale, 
provides an important contribution which is sensitive only to the 
longitudinal correlations of hadrons \cite{Strikman:2001gz}.
This is the contribution when two partons of the incident nucleon 
interact with partons belonging to two different nucleons in the nucleus, 
$\sigma_2$,
\begin{equation} 
\sigma_2 \;\; =\;\; \sigma_{\rm double}^{NN} \frac{A-1}{A}
\int d^2b \; T^2(b) \; \frac{f(x_1) f(x_2)}{f(x_1,x_2)} .
\end{equation}
The other term is the impulse approximation --- two partons of
the incoming nucleon interact with two partons of the same 
nucleon in the nucleus, $\sigma_1$, which is simply equal to 
$A$ times the cross section of double scattering in $pp$ collisions.
Thus, by measuring the ratio of $pA$ and $pp$ double scattering cross
sections we can determine $1/\sigma_{\rm eff}$ via the relation
\begin{equation} 
\frac{1}{\sigma_{\rm eff}} = 
\left[ \frac{\sigma_{\rm double}^{pA}}
{A  \sigma_{\rm double}^{pp} } -1\right] 
\; \left[\frac{1 - 1/A}{\int d^2b \; T^2(b)} \right] .
\end{equation}
This expression applies for $x_A \ge 0.03$, where nuclear effects in
the structure functions are small. Note that the experimental
measurement of the $A$--dependence will provide an independent 
test of this equation.
 
For the ratio of double to single scattering terms we find, for $A\ge 12$,
\begin{equation}
R \;\; = \;\; \frac{\sigma_2}{\sigma_1} \;\; \approx  \;\; 0.68 
\; \left(\frac{A}{12}\right)^{0.38} \; 
\frac{\sigma_{\rm eff}}{14 \; {\rm mb}} .
\end{equation}
Taking the CDF value of $\sigma_{\rm eff}\sim 14 \; {\rm mb}$, we obtain
$R\sim 3$ for $A\sim 200$. Thus, the separation of the two terms will be 
quite straightforward. Even in the case of deuteron--nucleus scattering, 
which was studied at RHIC recently, the contributions from two partons 
of one nucleon of the deuteron interacting with two different nucleons 
in the nucleus remains significant.It constitutes about 50\% of the 
cross section for $A \sim 200$. Hence we conclude that both 
measurements of $pA$ and $dA$ collisions will allow to measure of 
$\sigma_{\rm eff}$ if it is $\ge 5 \; {\rm mb}$, with $pA$ being 
a better option. Finally, if $\sigma_{\rm eff}$ will have been 
measured in $pA$ collisions, it will be possible to extract the
longitudinal two--parton distributions in a model independent way.

To summarize, we have demonstrated that future experiments will 
be able to measure independently the longitudinal and transverse 
two--parton distributions in the nucleon. With a detector of 
sufficiently large acceptance it would be possible to extend 
these studies even to the case of three parton correlations.
\section{The transverse spatial distribution of gluons
and diffractive Higgs production at LHC}
\begin{figure}
\begin{center}
\begin{tabular}{cl}
\includegraphics[width=9cm,height=6.3cm]{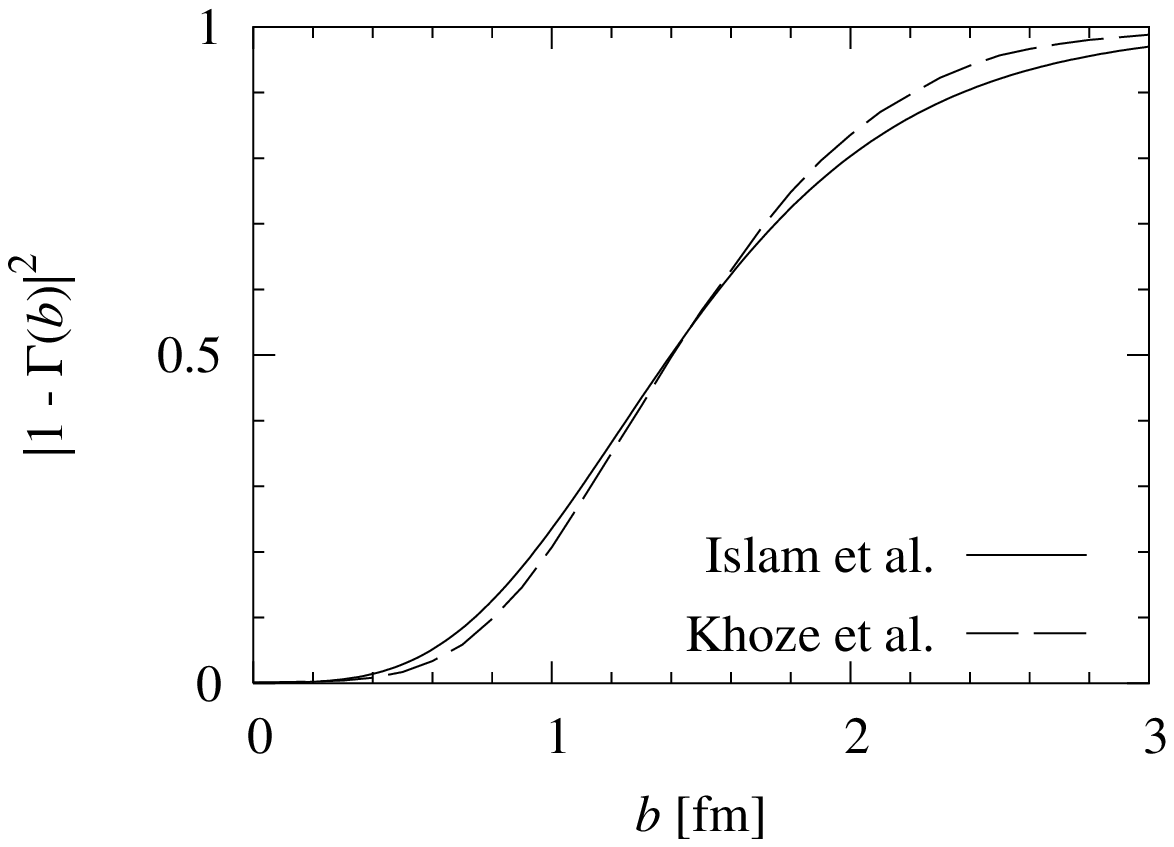}
& (a)
\\
\includegraphics[width=9cm,height=6.3cm]{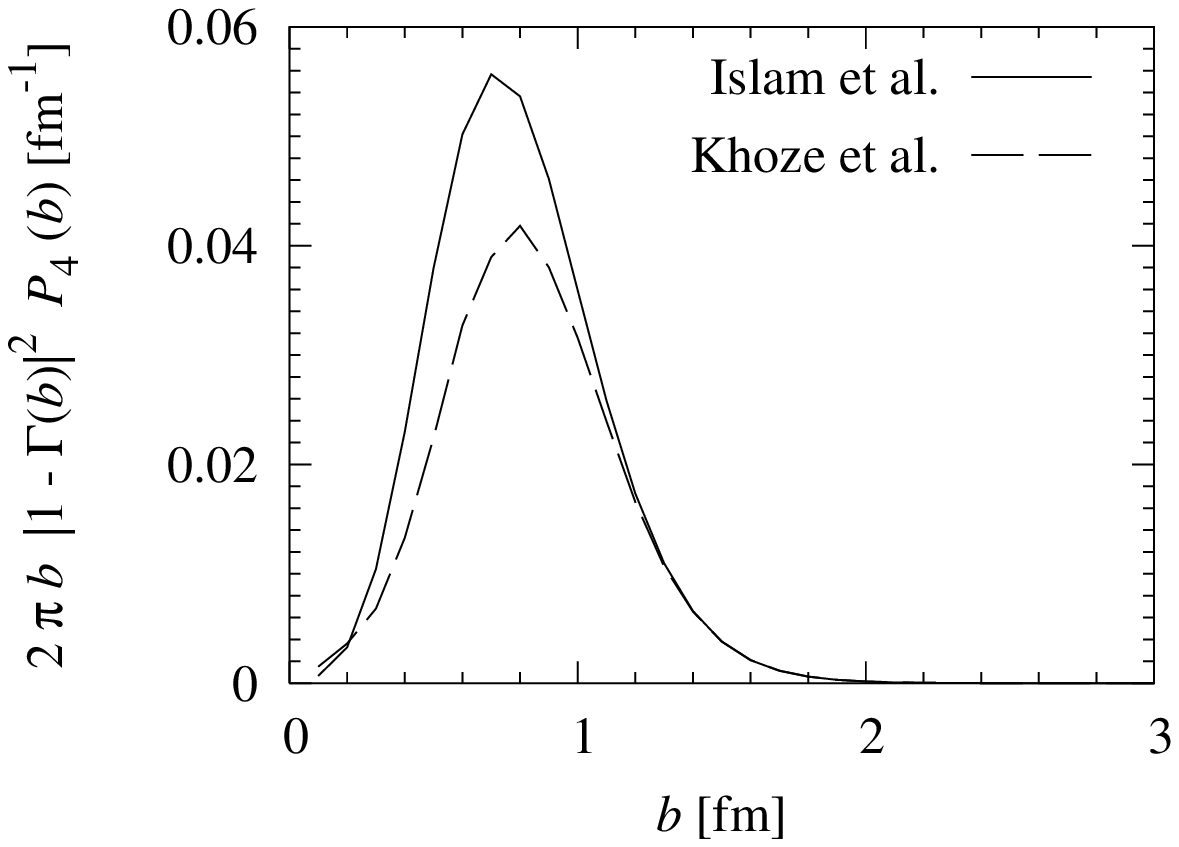}
& (b)
\end{tabular}
\end{center}
\caption[]{(a) The probability distribution for no inelastic interaction,
Eq.~(\ref{wnoin}), for $\sqrt{s} = 14\; {\rm TeV}$, as calculated with
the parametrizations of the $pp$ elastic amplitude of Islam et al.
\cite{Islam:2002au}, and Khoze et al. \cite{Khoze:2000wk}.
(b) The $b$--distribution for diffractive exclusive Higgs production, 
Eq.~(\ref{product}), as obtained with the dipole--type two--gluon 
form factor with $m_g^2 = 1\; {\rm GeV}^2$. Shown are the ``radial''
distributions including a factor $2\pi b$.}
\label{fig_wnoin}
\end{figure}

Exclusive diffractive production of Higgs bosons,
\begin{equation}
p + p \;\; \rightarrow \;\; p + \mbox{(gap)} + H + \mbox{(gap)} + p ,
\end{equation}
seems to be one of the promising candidates for the Higgs 
search at LHC; see Ref.~\cite{Kaidalov:2003fw} and references therein. 
From the point of view of strong interactions,
such processes involve a delicate interplay between ``hard'' and
``soft'' interactions, which leads to a characteristic dependence
of the cross section on the impact parameter of the $pp$ system, $b$. 
The heavy particle is produced in a hard partonic process 
(virtualities $\sim M_H^2$) involving the
exchange of two gluons between the nucleons --- one for the gluon--gluon 
fusion making the Higgs, the other for color neutralization. 
The impact parameter distribution for this process is described
by the function $P_4(b)$, see Eq.~(\ref{P_4}). In addition, 
the soft interactions between the two nucleons (viz.\ the spectator systems) 
have to conspire in such a way as not to fill the rapidity gaps
left open by the hard process. The probability for this can be
inferred from the amplitude of $pp$ elastic scattering in the 
impact parameter representation, $\Gamma (s, b)$. Namely,
\begin{equation}
|1 - \Gamma (s, b)|^2
\label{wnoin}
\end{equation}
is the probability for having no inelastic interaction in a
$pp$ collision with impact parameter $b$. This function can be
evaluated using available phenomenological parametrizations, 
which can be extrapolated to the LHC energy, $\sqrt{s} = 14 \; {\rm TeV}$. 
Fig.~\ref{fig_wnoin}a shows $|1 - \Gamma (s, b)|^2$ for the parametrization 
of Ref.\cite{Islam:2002au} and the multipomeron model of 
Ref.~\cite{Khoze:2000wk}. Both parametrizations indicate that
at this energy the nucleon is ``black'' [$|\Gamma(s, b)| \approx 1$]
for $b \leq 1\; {\rm fm}$. Fig.~\ref{fig_wnoin}b shows the
product 
\begin{equation}
|1 - \Gamma (s, b)|^2 P_4 (b) ,
\label{product}
\end{equation}
which governs the $b$--distribution of the cross section. 
The distribution is suppressed at small $b$ (because the probability 
for no inelastic interaction is small) as well as at large $b$
(because the overlap of the gluon distributions, $P_4(b)$, vanishes),
and is thus concentrated at intermediate values of the impact 
parameter, $b \sim 1\; {\rm fm}$. 
%
%

%
%
\begin{figure}
\begin{center}
\includegraphics[width=10cm,height=7cm]{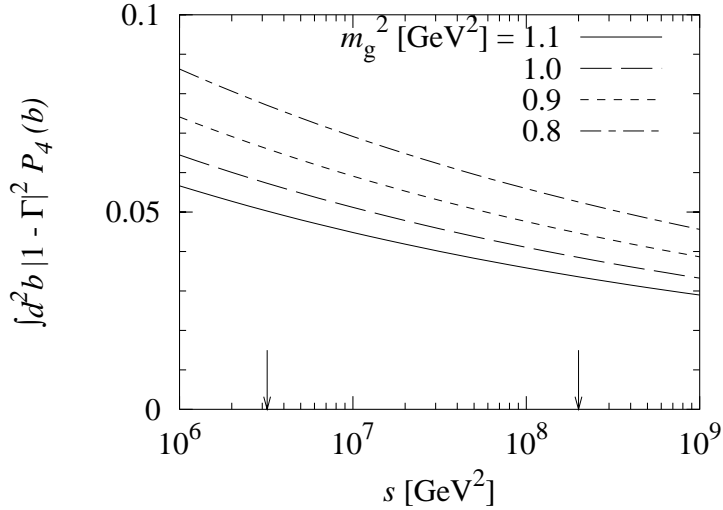}
\end{center}
\caption[]{The rapidity gap survival probability, $S^2$, 
Eq.~(\ref{surv}), obtained by integrating the product
$|1 - \Gamma (s, b)|^2 \; P_4 (b)$ shown in Fig.~\ref{fig_wnoin}
over impact parameters. Shown is the result as a function of $s$,
for various values of the mass parameter in the two--gluon form 
factor, $m_g^2$. The Tevatron and LHC energies are marked by arrows.}
\label{fig_surv}
\end{figure}
The integral of Eq.~(\ref{product}) defines the so-called rapidity
gap survival probability,
\begin{equation}
S^2 \;\; \equiv \;\; \int d^2 b \; |1 - \Gamma (s, b)|^2 \; P_4 (b) .
\label{surv}
\end{equation}
Fig.~\ref{fig_surv} shows the variation of this quantity with 
$s$ between Tevatron and LHC energies, for various values of the
dipole mass in the two--gluon form factor of the nucleon, Eq.~(\ref{dipole}).
The gap survival probability decreases with $s$ because the 
``black'' region in the proton grows with the collision energy.
Note that our results for $S^2$ are in agreement with those obtained by 
Khoze et al.\ \cite{Khoze:2000wk} in a multi--pomeron model,
given the uncertainty in the basic nucleon size parameter 
(denoted by $b$) in that approach, as well as with those 
reported by Maor et al.\ \cite{Maor}. In view of the different
theoretical input to these approaches this is very encouraging.
It is worth noting that, if we consider the production of an object 
of fixed mass at different energies, the values of $x_i$ decrease 
with the energy, corresponding 
to a smaller effective value of $m_g^2$ in Fig.~\ref{fig_surv}.
When this is taken into account, the actual drop of the survival probability 
with energy become much more modest.

We thank M. Ryskin for useful discussions.
This work was supported by U.S.\ Department of Energy Contract
DE-AC05-84ER40150, under which the Southeastern Universities Research 
Association (SURA) operates the Thomas Jefferson 
National Accelerator Facility.
L.~F.\ and M.~S.\ acknowledge support by the Binational Scientific 
Foundation. The research of M.~S. was supported by DOE.
\end{document}